\documentclass[11pt]{article}

\usepackage[utf8]{inputenc}
\usepackage{graphicx}
\usepackage{amsfonts}
\usepackage{amssymb}
\usepackage{amsmath}
\usepackage[mathscr]{eucal}
\usepackage{setspace}
\usepackage{booktabs}
\usepackage[shellescape,latex]{gmp}
\usempxpackage{amssymb}

\def\href#1#2{#2}   % Ignore hypertext links in UTPHYS.BST

\def\bb{\mathbb}
\def\sqr#1#2{{\vcenter{\hrule height.#2pt
   \hbox{\vrule width.#2pt height#1pt \kern#1pt
      \vrule width.#2pt}
   \hrule height.#2pt}}}

\def\bsqr#1#2{{\vrule width #1pt height#2pt}}
\def\bsquare{{\mathchoice\bsqr66\bsqr66\bsqr33\bsqr33}}
\def\badbreak{\penalty1000}

\def\Trs{{\mathop{\rm tr}}}		    % Trace (small)
\def\Trb{{\mathop{\rm Tr}}}		    % Trace (big)
\def\identity{{\bb I}}			    % Identity matrix
\def\diag{\mathop{\rm diag}}          % diagonal matrix
\def\R{{\bb R}}				    % Set of real numbers
\def\C{{\bb C}}				    % Set of complex numbers
\newcommand{\gfive}{\gamma_{5}}                  % gamma_5
\newcommand{\cF}{{\cal F}}                             % cal-F
\newcommand{\cO}{{\cal O}}                            % cal-O
\newcommand{\cS}{{\cal S}}                  % cal-S
\newcommand{\cT}{{\cal T}}                   % cal-T
\newcommand{\psibar}{{\bar\psi}}             % psi-bar
\def\vsR{{\scriptscriptstyle{\mathrm R}}}               % R subscript 
\def\vsI{{\scriptscriptstyle{\mathrm I}}}                  % I subscript 
\def\fir{{\scriptscriptstyle{\text{\rm IR}}}}                 % IR subscript 
\def\fuv{{\scriptscriptstyle{\text{\rm UV}}}}              % UV subscript 
\def\lm0{{\lambda_0}}                                             % IR scale 1
\def\cs{c_{\scriptscriptstyle {\mathrm S}}}               % c_S
\def\eff{{\scriptscriptstyle{\text{ef}}}}                       %  "eff" subscript/superscript
\def\ren{{\scriptscriptstyle{\mathrm R}}}
\def\scal{{\scriptscriptstyle{\mathrm S}}}

\makeatletter
\newcommand*{\smallrel}[2][.8]{%
\mathrel{\mathpalette{\smallrel@{#1}}{#2}}%
}
\newcommand*{\smallrel@}[3]{%
  \sbox0{$#2\vcenter{}$}%
  \dimen@=\ht0 %
  \raise\dimen@\hbox{%
    \scalebox{#1}{%
      \raise-\dimen@\hbox{$#2#3\m@th$}%
    }%
  }%
}
\makeatother

\usepackage{amsmath} % wonderful math package
\usepackage{dsfont}  % double strike font
\usepackage{bm}      % bold math

\def\beq{\begin{equation}}
\def\eeq{\end{equation}}
\def\beqs#1\eeqs{\beq\begin{split} #1 \end{split}\eeq}

\long\def\comment#1{}

\usepackage{microtype}
\usepackage[dvipsnames]{xcolor}
\usepackage[colorlinks=true,backref=false, linktocpage=true,
citecolor=PineGreen,urlcolor=PineGreen,linkcolor=PineGreen,pdfpagemode=UseOutlines]{hyperref}

\hypersetup{%
  bookmarksnumbered=true,
  pdftitle = {},
  pdfsubject = {},
  pdfauthor = {},
  pdfkeywords = {}
}

\begin{document}

\begin{center}
{\Large{\bf Gluon Condensate via Dirac Spectral Density:}} \\
\vspace*{.18in}
{\Large{\bf  IR Phase, Scale Anomaly and IR Decoupling}} \\
\vspace*{.24in}
{\large{Ivan Horv\'ath$^{a}$}} \\
\vspace*{.24in}
{\small Department of Physics and Astronomy, University of Kentucky, Lexington, KY 40503, USA} \\
\vspace*{.10in}
{\small Nuclear Physics Institute CAS, 25068 $\check{\text{R}}$e$\check{\text{z}}$ (Prague), Czech Republic} \\
\vspace*{.10in}
{\small Department of Physics, The George Washington University, Washington, DC 20052, USA} 

\vspace*{.15in}
{Oct 20 2025}

\end{center}

\vspace*{0.10in}

\begin{abstract}

Quark and gluon scalar densities, $\langle \bar{\psi} \psi \rangle$ and $\langle F^2 \rangle$, 
reflect the degree of scale-invariance violations in SU(N) gauge theories with fundamental
quarks. It is known that $\langle \bar{\psi} \psi \rangle$ can be usefully scale-decomposed 
via spectral density $\rho(\lambda)$ of Dirac modes. Here we give such formula for 
 $\langle F^2 \rangle$, which reveals that gluon condensate is a strictly UV quantity.  
For the recently-found IR phase~\cite{Alexandru:2015fxa, Alexandru:2019gdm}, where 
the infrared (IR) degrees of freedom separate out and become independent of 
the system's bulk, it implies that $\langle F^2 \rangle$ due to this IR part vanishes. 
Its glue thus doesn't contribute to scale anomaly of the entire system and is, 
in this sense, scale invariant consistently with the original claim. 
Associated formulas are used to define IR decoupling of glue, which may serve as 
an alternative indicator of IR phase transition. 
Using the simplest form of coherent lattice QCD, we express the effective action of full QCD 
entirely via Dirac spectral density.

%More generally, the formula reveals that gluon condensate is a strictly UV quantity and that 
%it arises due to the necessity of UV regularization.

\renewcommand{\thefootnote}{}
\footnotetext{\hspace*{-.65cm} 
\indent ${}^a${\tt ihorv2@g.uky.edu}
}
\end{abstract}

\vspace*{0.45in}

\vfill\eject

\noindent
{\bf 1.~Introduction.}
Quark and gluon scalar densities, namely the expectation values 
$\langle \bar{\psi} \psi \rangle$ and~$\langle F^2 \rangle$, play an important 
role in the dynamics of QCD. For example, in hadronic physics they are 
well-known as leading ``condensates"  entering the QCD sum 
rules~\cite{Shifman:1978bx}.  At a more basic level, they are connected 
to fundamental symmetries which makes them objects of prime interest 
for understanding the phases of nature's strong interactions and, more 
generally, the phases in a large class of theories where quarks and gluons 
interact in the same manner. For purposes of the present discussion, 
this class $\cT$ will consist of asymptotically-free vectorlike SU(3) 
gauge theories with fundamental quarks. Thus, $\cT$ involves theories 
with arbitrary number $N_f \!<\! 16.5$ of fundamental quark flavors 
with arbitrary masses $m_q$, $q=1,2,\ldots, N_f$ and at any temperature,  
including~$T\!=\!0$. 

Viewed through the lens of scalar densities, each theory from $\cT$ is 
characterized by $N_f \!+\!1$ values $\langle F^2 \rangle$,  
$\langle \bar{\psi} \psi_q \rangle$. But even their hypothetical full knowledge 
wouldn't provide us with understanding and classification of phases in the entire 
$\cT$. Indeed, the conventional symmetry-based apparatus for such analyses 
simply doesn't have a full reach within $\cT$. One useful but limited classification 
is usually done within the subclass of theories containing multiple massless 
flavors, where $\langle \bar{\psi} \psi \rangle$ indicates the spontaneous 
breakdown of the associated flavored chiral symmetry. Similar goes 
for considerations concerning the anomalous nature of flavor-singlet chiral 
symmetry. While these special circumstances and their implications are 
important for understanding certain aspects of low-energy ``real-world" QCD, 
they are in themselves unlikely to reveal the phase structure of the entire 
$\cT$. The situation is similar (but also different; see below) in case of dilations 
where full scale invariance requires quarks to be massless and temperature 
to be zero.

\begin{figure}[!b]
\vskip -0.10in
\centering
\hskip 0.00in
\includegraphics[scale=0.72]{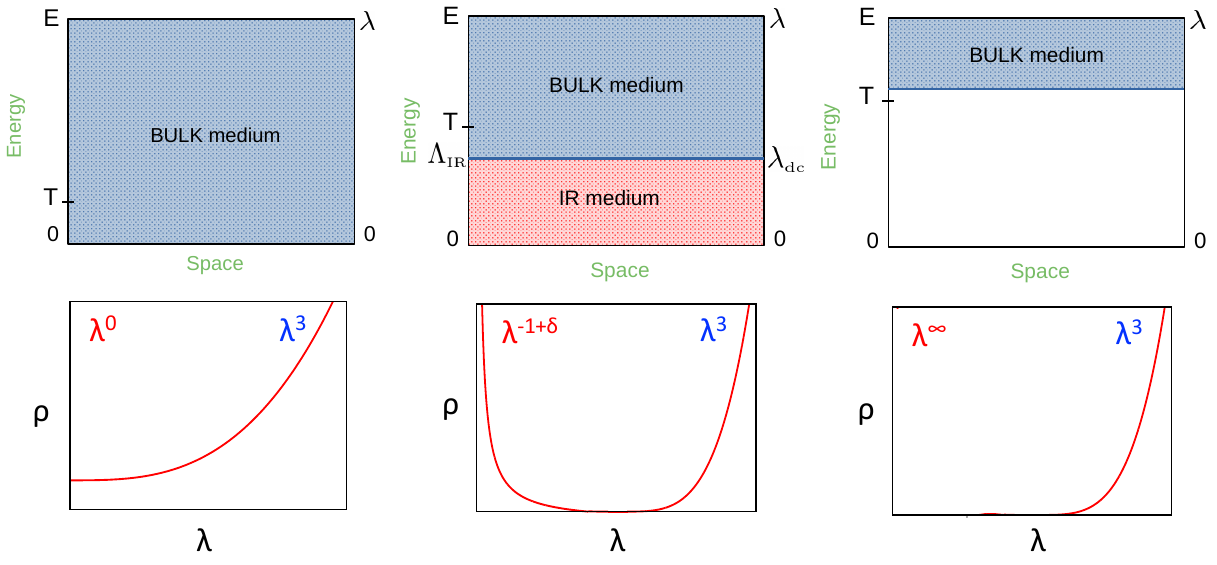}
\vskip -0.07in
\caption{Types of thermal states for theories in $\cT$ and schematics of their 
Dirac spectral densities $\rho(\lambda)$. Here $\lambda$ is the Dirac eigenvalue 
(scale) in the
continuum-like notation where $D\psi_\lambda = i \lambda \psi_\lambda$.
Left: B phase (standard confined phase) involves a single-component 
system with correlated parts. Its leading IR power behavior $\lambda^0$
includes cases when density is logarithmically divergent.
Middle: IR phase involves a multi-component system with IR separated
and decoupled from the bulk.
$\Lambda_\fir$ is the energy scale of IR-bulk separation and $\lambda_{dc}$ 
the associated Dirac scale. 
Right: the hypothetical UV phase describes a single-component system of 
weakly interacting quarks~and~gluons.
}
\vskip -0.20in
\label{fig:IR-bulk1}
\end{figure}

A dramatic departure from the traditional and purely symmetry-based considerations 
appeared in Refs.~\cite{Alexandru:2015fxa, Alexandru:2019gdm} which, 
together with refinements in Refs.~\cite{Alexandru:2021pap, Alexandru:2021xoi} 
and an additional state-of-the-art numerical evidence from 
Refs.~\cite{Meng:2023nxf, Alexandru:2023xho, Alexandru:2024tel}, led to 
the classification of phases in the entire $\cT$. This primarily stemmed from
the finding that systems in $\cT$ can partition their degrees of freedom and become 
multicomponent. More precisely, there is a dynamical regime within $\cT$ where 
deep infrared (IR) field fluctuations proliferate, separate out and decouple from 
the rest of finite-energy fundamental fields (bulk). Such IR-bulk separation, 
or property of being multi-component in general, doesn't depend on symmetries 
and can occur anywhere in $\cT$. The new regime became known as 
the {\em IR phase} of $\cT$ and was associated with restored scale 
invariance of glue in the IR component~\cite{Alexandru:2019gdm}. 

IR-bulk separation occurs in the Dirac space of theories in $\cT$, as 
shown schematically by Fig.~\ref{fig:IR-bulk1} in the setting of thermal 
transitions~\cite{Alexandru:2019gdm}. The theory that ``confines" at 
$T\!=\!0$~(left), such as pure-glue QCD or real-world QCD, describes 
a single-component system at low $T$ (blue bulk) and features field 
fluctuations correlated over any pair of Dirac ($\lambda$) and physical 
energy ($E$) scales. At certain temperature $T_\fir$ thermal fluctuations 
become strong enough to induce the creation of independent IR 
component (red medium) whose degrees of freedom do not participate 
in the makeup of hadrons (``partial deconfinement") and the IR phase 
ensues. Yet stronger disorder at $T \!>\! T_\fuv \!>\! T_\fir$ may inhibit 
deep-IR ($\lambda \!\neq\! 0$) degrees of freedom, with the system 
becoming single-component again, possibly as a ``deconfined" perturbative 
bulk. It is worth noting that, unlike the first two cases, this classic scenario 
for weakly-coupled quark-gluon plasma has not yet been clearly identified 
in QCD and its existence is uncertain. Schematics of Dirac spectral 
density $\rho(\lambda)$ in Fig.~\ref{fig:IR-bulk1} conveys how quark 
degrees of freedom are distributed across scales in the three regimes.

Since the separation of components is scale-based it is desirable, if 
at all possible, to think of scalar densities as composed of 
scale-dependent parts. Note that spectral density already partitions 
degrees of freedom by scale. This aspect was thus innately present 
in the arguments of original~works on IR 
phase~\cite{Alexandru:2015fxa, Alexandru:2019gdm}, including 
arguments for scale invariance of the IR component. However, 
more formal treatment does need a scale decomposition of scalar 
densities since departures from scale invariance due to the glue field 
and the quark field of mass $m$ are quantified by 
(scale anomaly)~\cite{Nielsen:1975, Collins:1977a, Nielsen:1977}
\begin{equation}
    T_{\mu\mu} \,=\, \frac{\beta(g)}{2g} \, 
    \langle F^2 \rangle  \,+\, \bigl( 1+ \gamma_m(g) \bigr) 
    \,m \langle \bar{\psi}\psi \rangle
    \label{eq:007}
\end{equation}
and only scale decompositions of $\langle F^2 \rangle$ and 
$m \langle \bar{\psi}\psi \rangle$ can isolate the contribution of 
a component.\footnote{Notation here is standard, with $T$ 
the energy-momentum tensor and $\beta$, $\gamma_m$ 
the conventional RG functions.} To that end, we point out here 
the usefulness of the following expressions
\begin{equation}
    -m \langle \psibar\psi \rangle  = 
    m \int_{\R^2[\C]} d\cS \;  \frac{1}{\lambda+m} \;\, \rho_s^\eff(\lambda)    
    \qquad\qquad\;
    \langle \,F^2\, \rangle = \frac{a}{\cs} \, 
    \int_{\R^2[\C]} d\cS \;  \lambda \;\, \rho_s^\eff(\lambda) 
    \quad
    \label{eq:017}      
\end{equation}
with $\lambda \!=\! \lambda_\vsR \!+\!  i\lambda_\vsI$ a complex 
eigenvalue of lattice Dirac operator $D$ defining quark dynamics 
in a regularized theory (UV cutoff $1/a$; IR cutoff $1/L$). This $D$ 
is chosen such that the associated dimensionless constant $\cs$ 
is non-zero (see below). Quantity 
$\rho_s(\lambda) \!=\! \langle n(\lambda, d\cS)\rangle /(V_4 \,d\cS)$
is the {\em surface spectral density} of $D$, with $n(\lambda, d\cS)$ 
the number of eigenvalues in the area 
$d\cS \!=\! d\lambda_\vsR d\lambda_\vsI$ around $\lambda$, and 
$V_4 =L^3/T$.\footnote{Denoting the integration domain $\R^2[\C]$ 
conveys that components of $\R^2$ have meaning in the complex 
plane $\C$.} 
Dimension of $\rho_s(\lambda)$ is $a^{-2}$ instead of $a^{-3}$ 
in the standard definition.
Effective density $\rho_s^\eff \equiv \rho_s \!-\! \rho_{s0}$ subtracts 
that of the free field.\footnote{We will see that in the glue case 
the free-field subtraction is necessary to merely define the operator. 
In the quark case it removes the leading UV divergence at non-zero 
$m$, which keeps the two expressions symmetric in this sense.} 
Note that  $\rho_s^\eff = \rho_s^\eff(\lambda,a,L)$ and 
$\int_{\R^2[\C]} d\cS \; \rho_s^\eff(\lambda)=0$.

Albeit not in this form, the quark relation in \eqref{eq:017} 
appeared in considerations leading to Banks-Casher 
relation~\cite{Banks:1979yr}. The glue expression is new and 
exploits the ideas put forward in 
Refs.~\cite{Horvath:2006az, Horvath:2006md, Alexandru:2008fu}.
Together they expose, in an explicit way, the key difference 
between the two scalar densities in terms of their scale makup: 
while the kinematic factor $1/(\lambda + m)$ enhances IR and 
suppresses UV in quark case, its counterpart $\lambda$ does 
the exact opposite in glue case. In fact, the glue expression invokes 
a surprising possibility that gluon condensate is an entirely UV 
quantity. 
We will show how Eqs.~\eqref{eq:017} clarify the proposed 
relation of IR phase to scale invariance~\cite{Alexandru:2015fxa, 
Alexandru:2019gdm}, and how to use the associated formulas 
to characterize its IR-bulk decoupling. We make all these results 
solid by writing the corresponding expressions 
(Eqs.~\eqref{eq:051} and \eqref{eq:061}) for the important 
case of the overlap Dirac operator.

%We will point out that equations in \eqref{eq:017} allow for alternative 
%definition of IR phase and clarify its proposed relation to scale 
%invariance~\cite{Alexandru:2015fxa, Alexandru:2019gdm}. They 
%also expose, in a clear manner, the key difference between the two 
%scalar densities in terms of their scale content: while the kinematic 
%factor $1/(\lambda + m)$ enhances IR and suppresses UV in quark 
%case, its counterpart $\lambda$ does the exact opposite in glue 
%case. In fact, the glue expression invokes a surprising possibility 
%that gluon condensate is an entirely UV quantity. We will make 
%such conclusions solid by writing relations \eqref{eq:017} for 
%the important case of the overlap Dirac operator. 

\smallskip
\noindent
{\bf 2.~Gluon Condensate.}
We now derive the expression for gluon condensate in \eqref{eq:017}.
To that end, consider the Euclidean lattice setup for theories in $\cT$, 
consisting of $N_s^3 \times N_\tau$ sites of a hypercubic lattice with 
UV cutoff $1/a$, IR cutoff $1/L$ ($L \!=\! N_s a$), and temperature 
$T$ ($1/T \!=\! N_\tau a$). Lattice spacing $a$ is controlled by 
the gauge coupling $g$ but this relationship can be kept implicit here. 
Let $U \!\equiv\! \{U_\mu(x)\}$ be a configuration of SU(3) gauge field 
and $D \!=\! D(U)$ a lattice Dirac operator. The latter is explicitly 
assumed to be local, gauge covariant and to respect hypercubic 
symmetries. Local {\em gauge operators} of definite space-time 
transformation properties can then be constructed from $D$ via 
suitable tracing 
operations~\cite{Horvath:2006az, Horvath:2006md}. In case of 
$F ^2 (x) \equiv  \Trs_c \, F_{\mu\nu}F_{\mu\nu}(x)$ one starts 
from~\cite{Horvath:2006az, Horvath:2006md, Alexandru:2008fu}
%\footnote{Note that, in Refs.~\cite{Horvath:2006az, Horvath:2006md}, 
%$D$ denoted Dirac operator  in lattice units which explains the shift 
%in powers of $a$.} 
\begin{equation}
     \Trs_{cs} \hat{D}_{x,x}(U) - \Trs_{cs} \hat{D}_{x,x}(\identity)  \,=\,  
     \cs \,a^4 \,\Trs_c\, F_{\mu\nu}F_{\mu\nu}(x,A)  + \cO(a^6)
     \quad
     \label{eq:011} 
\end{equation}
for transcription $U$ of a classical continuum gauge field $A$ onto 
the hypercubic lattice with~spacing~$a$, and 
$F_{\mu\nu}=\partial_\mu A_\nu - \partial_\nu A_\mu + i \,[A_\mu,A_\nu]$.
Here $\identity$ denotes the free field $U_\mu(x) \!\equiv\! \diag\{1,1,1\}$,
$\Trs_c$ is the trace in color while $\Trs_{cs}$ in color-spin, and $\cs$ is 
a constant. Note that $\hat{D} = a D$ is dimensionless but $F_{\mu\nu}$ 
has its physical dimension. 
[$\,$In Refs.~\cite{Horvath:2006az, Horvath:2006md, Alexandru:2008fu}
$D$ denotes operator in lattice units.]~Eq.~\eqref{eq:011} implies that, if 
$\cs \!\ne\! 0$, quantum operator $F^2(x)$ can be defined via indicated 
Dirac matrix elements,~namely  
\begin{equation}
     F^2(x,U)  \equiv \frac{1}{\cs \,a^3} \,    
     \Trs_{cs} \Bigl[ D_{x,x}(U) - D_{x,x}(\identity) \Bigr] 
     \quad\, \longrightarrow \quad\,\,
     \langle \,F^2\, \rangle = \frac{a}{\cs} \,\frac{T}{L^3} \,
     \Bigl\langle \, \Trb \Bigl[ D(U) - D(\identity) \Bigr] \,\Bigr\rangle \;
     \label{eq:021}   
\end{equation}
where ``$\Trb$" denotes the full trace of Dirac matrix. When performing 
the QCD average, we replaced $F^2(x)$ with 
$(\sum_yF^2(y))/(N_s^3 N_\tau)$, permitted by virtue of hypercubic 
translation invariance. For general Ginsparg-Wilson 
operators~\cite{Ginsparg:1982}, non-zero $\cs$ is expected due to their 
non-ultralocality~\cite{Horvath:1998cm, Horvath:1999bk, Horvath:2000az}.
In case of overlap Dirac operators~\cite{Neuberger:1997fp} this was shown 
via explicit computation in Ref.~\cite{Alexandru:2008fu}.

Unlike in the continuum, eigenvalues $\lambda_j$ of lattice $D$ are not 
purely imaginary and details of the spectrum vary in different formulations. 
To cover all possibilities, we introduced in Sec.~1 the surface spectral 
density $\rho_{s}(\lambda)$ of eigenvalues in complex plane. This density 
is usefully represented as
$\rho_s(\lambda, U) \!=\! 
\sum_j  \delta(\lambda_\vsR - \lambda_\vsR^j) \,
             \delta(\lambda_\vsI - \lambda_\vsI^j)/V_4$ with
$\lambda \!=\! \lambda_\vsR \!+\! i \lambda_\vsI$, $\lambda^j=\lambda^j(U)$ 
and $V_4 = L^3/T$. Then
\begin{equation}
     \langle \, \Trb \, D \,\rangle \,=\, 
     V_4 \int_{\R^2[\C]} d\cS \, \lambda \; \rho_s(\lambda)  
     \qquad , \qquad
     \rho_s(\lambda) \equiv  \langle\, \rho_s(\lambda, U) \,\rangle
     \label{eq:031}   
\end{equation}
which consequently turns the expression for gluon condensate in 
Eq.~\eqref{eq:021} into one in Eq.~\eqref{eq:017} as~desired.

\smallskip
\noindent
{\bf 3.~Gluon Condensate with Overlap Operators.} To analyze 
the implications of general Eqs.~\eqref{eq:017}, it is desirable 
to examine them for operators $D$ that mimic continuum 
features to the largest extent possible. To that end, first note 
that neither quark nor gluon lattice condensate is guaranteed 
to be real-valued a priori. A natural way to ensure this is to impose 
$\gfive$-Hermiticity, namely $\gfive D \gfive = D^\dagger$, which 
forces eigenvalues to appear in complex-conjugate pairs, and 
physically amounts to continuum-like treatment of antiparticles. 
We will thus assume that $D$ is $\gfive$-Hermitian from~now~on.

It is also desirable to ensure continuum-like chiral properties 
on the lattice. At present, this amounts to using Ginsparg-Wilson 
(GW) Dirac operators~\cite{Ginsparg:1982}.\footnote{See e.g. 
Refs.~\cite{Horvath:1999bk, Horvath:2000az} for full specification
of the class.} Among them, the 1-parameter family of 
{\em overlap operators} $D$ based on Wilson-Dirac 
matrix~\cite{Neuberger:1997fp} has been studied most extensively.
They are given by 
\begin{equation}
   \frac{a}{\Delta} \, D  \,=\, 
   1 + \frac{\hat{D}_W- \Delta}{\sqrt{\bigl( \hat{D}_W - \Delta \bigr)^\dagger
                                                        \bigl( \hat{D}_W - \Delta \bigr) }}
    \quad\quad , \quad\quad
    \Delta \in (0,2)
     \label{eq:041}       
\end{equation}
where $\hat{D}_W$ is the dimensionless massless Wilson-Dirac operator. 
In Ref.~\cite{Alexandru:2008fu} it was shown that 
$\cs \!=\! \cs(\Delta)\!\neq\! 0$ for the above $D\!=\!D(\Delta)$. They 
can thus be used to express gluon condensate~via~Eq.~\eqref{eq:017}.

To that end, note that the $\gfive$-Hermitian spectrum of $D$ traces 
the circle of radius $\Delta$ centered at $(\Delta,0)$ so that 
the eigenvalues $\lambda_\vsR + i \lambda_\vsI$ satisfy
$a(\lambda_\vsR^2 + \lambda_\vsI^2) = 2\lambda_\vsR \Delta$. 
The expression involving spectral density along the circle is obtained via
substituting $\rho_s(\sigma \cos\varphi, \sigma \sin\varphi)  \!=\! 
\rho(\sigma)\delta(\varphi-\cos^{-1}(a\sigma/2\Delta))/\sigma$ upon
transfer to polar coordinates in Eq.~\eqref{eq:017}. We obtain
\begin{equation}
     \langle \,F^2\, \rangle_{a,L} \;=\; \frac{a^2}{\cs\Delta} \,
      \int_{0}^{\bigl(\frac{2\Delta}{a}\bigr)^-} \! d\sigma \, \sigma^2  \, 
      \rho^\eff(\sigma, a, L)      \;\, + \;\,\,
      T \, \frac{\langle n_0 \rangle_{a,L}}{L^3} \, \frac{2\Delta}{\cs}
      \qquad
      \label{eq:051}   
\end{equation}
where $\rho^\eff \!=\! \rho - \rho_0$ has been defined previously and 
$\langle n_0 \rangle$ is the average number of exact zeromodes equal
to the number of modes with real eigenvalue $2\Delta/a$. While the former 
do not contribute to $\langle F^2 \rangle$, the discrete contribution 
of the latter was separated out. In the continuum limit, energy-like variable 
$\sigma$ (magnitude of lattice eigenvalue) coincides with $\lambda$ of 
the continuum Euclidean formulation where eigenvalues are parametrized 
by $i \lambda$. Thus, $\rho(\sigma)$ is associated with the upper branch 
($\lambda \!\ge\! 0$) of the continuum density. 
Analogous derivation for the quark expression in~\eqref{eq:017} leads to
\begin{equation} 
   -m \langle \psibar\psi \rangle_{a,L}  = 
    \int_{0^+}^{\bigl(\frac{2\Delta}{a}\bigr)^-} \!d\sigma \;
    \frac{m^2 + m^2 + a \sigma^2 m / \Delta}
           {m^2 + \sigma^2  + a \sigma^2 m / \Delta } \;\, 
    \rho^\eff(\sigma, a, L)    \;\, + \;\,\,
    T \, \frac{\langle n_0 \rangle_{a,L}}{L^3} \,
    \Bigl( 1 + \frac{am}{2\Delta + am} \Bigr)
    \quad      
    \label{eq:061}       
\end{equation}
Here all real modes contribute and are separated out in the second term 
of the expression. We emphasize that Eqs.~\eqref{eq:051} and 
\eqref{eq:061} are fully regularized scale decompositions of these 
quantities. 

\smallskip
\noindent
{\bf 4.~The Uses: Gluon Condensate as a UV Quantity.}
The above implies that, in a well-defined sense, gluon condensate in 
QCD is a UV quantity. The underlying logic is that the contribution of 
Dirac scales up to any finite renormalized value $\lambda_\ren$ 
vanishes in the continuum limit. Indeed, writing 
the associated integral in Eq.~\eqref{eq:051} via renormalized 
quantities $\sigma_\ren = \sigma/Z_\scal$, $m_\ren = m/Z_\scal$, 
$\rho_\ren(\sigma_\ren) \!=\! 
Z_\scal \,\rho(Z_\scal \sigma_\ren)$~\cite{DelDebbio:2005qa, Giusti:2008vb} 
we obtain 
\begin{equation}
     \langle \,F^2\, \rangle_{L} \bigl[ \lambda_\ren \bigr] 
     \;=\; \lim_{a \to 0}  \, \frac{a^2 Z_S^2}{\cs\Delta} \,
      \int_{0}^{\lambda_\ren} \! d\sigma_\ren \, 
      \sigma_\ren^2  \, 
      \rho_\ren^\eff(\sigma_\ren, a, L)
      \;=\;0      
      \quad\; , \quad\;  \forall \;\, 0 < \lambda_\ren < \infty
      \quad
      \label{eq:071}   
\end{equation}
where the notation $\langle \,F^2\, \rangle [ \lambda_\ren ]$ means 
discarding all bare Dirac scales whose renormalized counterpart exceeds
$\lambda_\ren$.  
The limit is zero since $\rho_\ren^\eff$ as well as the associated 
integral have a well-defined continuum limit at fixed $L$, and since 
the $a$-dependence of $Z_\scal$ is at most logarithmic. Note also that 
the real modes, separated in Eq.~\eqref{eq:051}, contribute at the scale 
of UV cutoff and thus do not enter Eq.~\eqref{eq:071}.

Two additional remarks are worth making. Firstly, the above argument 
was made in the context of a finite system of size $L$. Regarding 
the infinite system, the same reasoning and conclusion applies for the order 
of limits $\lim_{L \to \infty}\lim_{a \to 0}$. For the reversed order, the first limit  
$\lim_{L \to \infty}$ is finite but $a$-dependent, and could in principle lead
to such an UV divergence of the integral in \eqref{eq:071} that would produce 
a non-zero contribution to gluon condensate. However, such UV divergence 
is unlikely to occur since it would have to be generated by IR Dirac modes. 
We thus propose that vanishing of any IR contribution to gluon condensate 
holds for both ways of removing cutoffs.

Secondly, note that the situation is very different for the quark part of trace 
anomaly. At finite fixed $L$ we have from Eq.~\eqref{eq:061} upon taking 
the continuum limit
\begin{equation} 
   -m \langle \psibar\psi \rangle_L \bigl[ \lambda_\ren \bigr]          \,=\,
   2  \int_{0^+}^{\lambda_\ren} \! d\sigma_\ren \, 
    \frac{1}{1 + \sigma_\ren^2/m_\ren^2} 
    \;\rho^\eff(\sigma_\ren, L)   \;\,+\;\, 
    T \, \frac{\langle n_0 \rangle_{L}}{L^3} 
    \quad      
    \label{eq:081}       
\end{equation}
where the second term is due to zero modes. The IR contribution is 
clearly non-zero with the second term vanishing in the $L \to \infty$ 
limit. The above form also shows that the quark contribution to trace 
anomaly becomes a strictly IR quantity in the chiral limit, which is 
familiar from considerations leading to Banks-Casher 
relation~\cite{Banks:1979yr}. 

\begin{figure}[t]
	\vskip -0.1in     
	\centering   
	\includegraphics[width=0.36\linewidth]{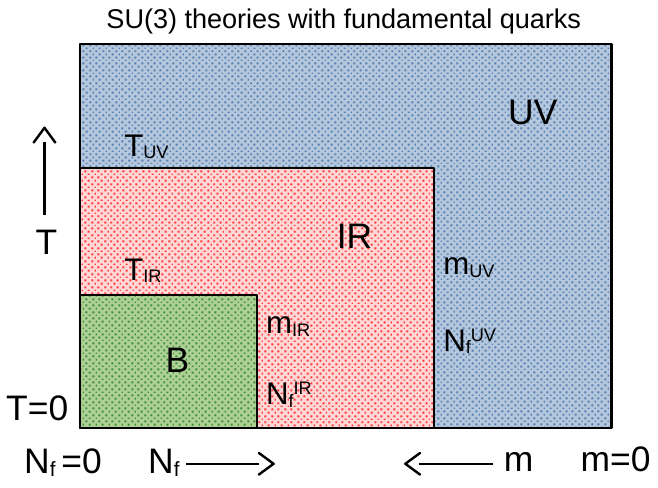}
	\hskip 0.35in
	\includegraphics[width=0.36\linewidth]{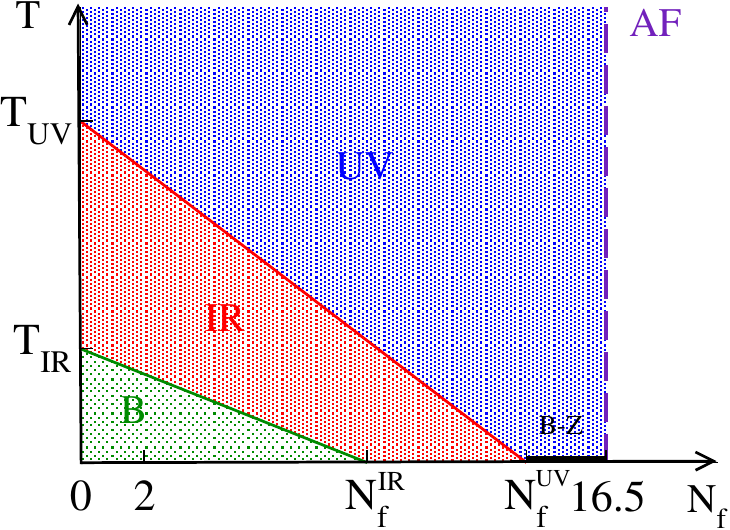}
	\caption{\label{fig:set_cT} Left: schematic view of phases in set ${\cal T}$ 
	based on a degree of deep-IR degrees of freedom proliferation and their 
	scale invariance. Direction of arrows for parameters indicates the direction
	of possible phase changes along the chain 
	$\text{B} \!\to\! \text{IR} \!\to\! \text{UV}$~\cite{Alexandru:2019gdm}.
	Right: the case of near-massless quarks with Banks-Zaks (B-Z) regime
	and the asymptotic freedom (AF) boundary indicated.}  
	\vskip -0.1in
\end{figure}

\smallskip
\noindent
{\bf 5.~The Uses: IR Scale Invariance in IR Phase.}
In the original work~\cite{Alexandru:2019gdm}, IR phase was defined 
by the negative power in IR behavior of Dirac spectral density, and
thus by a power-law enhancement of deep-IR degrees of freedom. 
The associated classification of phases in set $\cT$ is
\begin{equation}
     \rho(\lambda)   \propto   \lambda^p   \; , \;    
     \lambda \to 0    \quad\;\, \Longrightarrow \quad\;\,
     \text{phase}  \;\,=\;\,
     \text{B}  \;\,  \text{if}  \;\;  p=0 
     \;\;\; , \;\;\;\text{IR}   \;\,  \text{if} \;\;  p<0  
     \;\;\; , \;\;\;\text{UV}  \;\,  \text{if} \;\;  p>0 
     \;\;\;      
     \label{eq:091}
\end{equation}
where continuum-like notation ($\sigma \to \lambda$) was used but 
the classification is already well-defined on the lattice. In IR phase, $p$ 
is usefully written as $p \!=\! -1 \!+\! \delta$ because, at least in thermal 
cases, $\delta \!>\! 0$ is small and may vanish in the continuum 
limit~\cite{Alexandru:2019gdm}. 
The ``B" in B phase refers to broken IR scale invariance of glue: 
the standard IR characteristic of ``confined phase". Here spectral density 
is expected to logarithmically diverge away from the chiral 
limit~\cite{Osborn:1998qb, Alexandru:2019gdm}, and thus $p\!=\!0$. 
In not-yet-observed UV phase, IR degrees of freedom are power-law 
suppressed with the power possibly infinite if spectral density is zero in 
deep IR. The associated phase structure of $\cT$ is shown in 
Fig.~\ref{fig:set_cT} (left).

Two aspects~\cite{Alexandru:2019gdm} of the new IR phase are especially 
relevant for the present discussion.\footnote{See Ref.~\cite{Horvath:2025ypt} 
for more detailed discussion of features that make $p \!<\! 0$ regime a truly 
distinct phase.} 
{\bf (i)}  Due to $p \!<\! 0$ in IR and $p \!=\! 3$ in UV (asymptotic freedom), with 
an intervening regime of severe mode depletion, $\rho(\lambda)$ of theories 
in IR phase exhibits a bimodal structure with the IR part separated from the bulk 
in scale/energy. This invoked the conjecture of IR-bulk decoupling, namely that 
the IR part is independent from the bulk and acts as an autonomous subsystem. 
{\bf (ii)} Due to the near-pure negative-power behavior of $\rho$ and 
the connection to conformal 
window~\cite{Alexandru:2019gdm, Alexandru:2014paa, Alexandru:2014zna}
(see Fig.~\ref{fig:set_cT} (right)), it was conjectured that {\em glue} of the IR 
part is scale invariant, at least asymptotically. Thus, although defined via 
Dirac spectral properties, the physics behind the above structure of $\cT$ 
was proposed to be driven by glue and interpreted in glue 
terms~\cite{Alexandru:2015fxa, Alexandru:2019gdm}. In that vein, the new 
Eq.~\eqref{eq:051} is perhaps the purest expression of the implied 
connection and allows us to express the key properties of IR phase, such as  
{\bf (i)} and {\bf (ii)}, in more standard field-theoretic terms. 

To that end, it is important to realize that {\bf (i)} and {\bf (ii)} are in fact 
connected~\cite{Alexandru:2019gdm,Alexandru:2021pap}. For example,
while the IR-bulk separation is very suggestive of IR-bulk decoupling, it is not
sufficient for the claim. But  {\bf (ii)} offers a dynamical reason for it which is easy to see  
in thermal IR phase at $N_f\!=\!0$. Indeed, assume that, upon entering 
the IR phase, this theory of glue becomes scale invariant below some energy 
$\Lambda_\fir < T$.  The gauge coupling then stops running at that scale, 
which requires non-analyticities in the internal structure of the theory. 
These non-analyticities can then also facilitate the IR-bulk decoupling. 
Conversely, non-analyticities generated by breaking-off the IR into 
a physically distinct component can make scale invariance in IR 
possible.\footnote{Thermal onset of these 
internal non-analyticities is predicted to occur at $T\!=\!T_\fir$ of IR 
phase transition~\cite{Alexandru:2021pap, Alexandru:2021xoi} because they 
induce non-analytic $T$-dependence of physical observables at that point.}

There are at least two aspects of scale invariance to study in these IR 
phase circumstances. The first one isolates the IR component as a field 
system, bringing $\Lambda_\fir \rightarrow \infty$ by virtue of an overall
rescaling, and aims to study scale-invariant field theory so defined, e.g. 
in the strongly-coupled part of conformal window. 
The second one views the multi-component system as a whole 
and seeks to understand the role its IR part plays in violations of scale 
invariance. Focusing on this second aspect of {\bf (ii)}, let 
$\lambda_{dc}=\lambda_{dc}(a)$ be the scale marking the IR-bulk 
boundary in the Dirac spectrum (see Fig.~\ref{fig:IR-bulk1}). 
The contribution $\langle F^2 \rangle_\fir$ of the IR component (IR medium) 
to full $\langle F^2 \rangle$ is 
\begin{equation}
     \langle \,F^2\, \rangle_{\fir} 
     \;\equiv\;   \, \frac{a^2}{\cs\Delta} \,
      \int_{0}^{\lambda_{dc}(a)} \! d\sigma \, 
      \sigma^2  \, \rho^\eff(\sigma, a)
      \quad \longrightarrow \quad\,  0 
      \quad\;\, \text{for} \quad\;\, a \;\longrightarrow\; 0 
      \quad
      \label{eq:101}   
\end{equation}
Its approach to zero in the continuum limit ensues due to obvious 
integrability at any finite $a$, and because $\lambda_{dc}(a)$ varies at most 
logarithmically in the vicinity of $a \!=\! 0$. Hence, the IR component of 
the system in IR phase doesn't contribute to scale anomaly and is, from this 
point of view, scale invariant. The above argument is generic for IR phase 
of theories in $\cT$, and doesn't depend on whether $L$ is kept fixed 
(finite) or taken to infinity at each $a$.

\smallskip
\noindent
{\bf 6.~The Uses: IR-Bulk Decoupling in IR Phase.}
Aspects of the present analysis, and those of 
Refs.~\cite{Horvath:2006az, Horvath:2006md, Alexandru:2008fu}, 
can also be used to study IR-bulk decoupling in IR phase 
(point {\bf (i)} in Sec.~5).
The chief idea is that ``decoupling" is identified with ``decorrelation"  
which turns such parts into independent subsystems. Here we analyze 
the glue part of the system which is our focus in this work. The full 
account of spectral correlations and decoupling will be given elsewhere.

Since the IR-bulk separation is based on Dirac scales, we need 
the notion of correlation among~such scale-based parts. This
requires analogues of expressions in previous sections but for 
a given gauge background $U$. To that end, let's consider 
the action-like dimensionless construct $\cF^2(U)$
\begin{equation}
     \cF^2(U) \,\equiv\, a^4 \sum_x F^2(x,U) \,=\,   
     \frac{a}{\cs} \, \Trb \Bigl[ D(U) - D(\identity) \Bigr]  \,=\,
     V_4 \, \frac{a}{\cs} \int_{\R^2[\C]} d\cS \, \lambda \; \rho_s^\eff(\lambda,U)  
     \quad
     \label{eq:111}        
\end{equation}
written here for general $D(U)$ with $\cs \!\neq\! 0$, and where 
Eqs. \eqref{eq:021} and \eqref{eq:031} and their notation were used. 
For overlap Dirac operators we then have a specific expression 
(analogous to \eqref{eq:051}), which allows to quantify the contribution 
to $\cF^2(U)$ from range of lattice Dirac scales 
$\sigma$ ($0 \le \eta_1 \le \sigma \le \eta_2 < 2\Delta/a$)~as
\begin{equation}
     {\cal F}^2(U, \eta_1, \eta_2)  
     \,\equiv\,  \frac{V_4 a^2}{\cs\Delta} \,
     \int_{\eta_1}^{\eta_2} \! d\sigma \, 
     \sigma^2  \, \rho^\eff(\sigma,U)
     \;=\; \frac{a^2}{\cs\Delta} \, 
     \sum_{\eta_1 < \sigma_j \le \eta_2} 
     \Bigl[ \, \sigma_j^2(U)  - \sigma_j^2(\identity) \Bigr]
     \quad    
      \label{eq:121}        
\end{equation}
where the filter only allows values  $\sigma_j(U)$ and (separately) 
$\sigma_j(\identity)$ in the specified range. Note that the formula 
for given non-negative $\eta_1$, $\eta_2$ in fact combines (equal) 
contributions from both upper and lower branch of the Dirac 
spectrum. It is fully regularized and allows for defining correlations 
amongst different ``parts" of the system with ``partitions" based 
on Dirac spectra.

To formulate IR-Bulk decoupling of glue in IR phase, it is practical 
to make two preparatory steps. First, it is convenient to replace 
the variable $\sigma$ with $t \!=\! \sigma/m$, where $m$ may be 
e.g. the smallest non-zero quark mass in the theory, because 
it doesn't get renormalized~\cite{DelDebbio:2005qa, Giusti:2008vb}. 
Secondly, rather than the usual connected correlation 
$\langle \cO_1 \cO_2 \rangle_\text{c}  \! \equiv \! 
\langle \cO_1 \cO_2 \rangle - \langle \cO_1\rangle \langle \cO_2 \rangle$, 
we consider the {\em normalized}
connected Pearson correlation $\langle \ldots \rangle_\text{nc}$
(covariance divided by standard deviations),~namely
\begin{equation}
    \langle \cO_1 \cO_2 \rangle_\text{nc} \!\equiv\!
    \frac{\langle \cO_1 \cO_2 \rangle_\text{c}}
    {\sqrt{\langle \cO_1 \cO_1 \rangle_\text{c}  
              \langle \cO_2 \cO_2 \rangle_\text{c} }} 
      \label{eq:131}        
\end{equation}
which scales out magnitudes of correlated observables and thus 
expresses pure correlation. 

Let $t_\text{ref}$ be an arbitrary but fixed Dirac scale inside the bulk 
of the target continuum theory in IR phase defined by 
Eq.~\eqref{eq:091}.  We note in passing that the boundary between 
IR and bulk in IR phase may be defined by the existence of a point 
$t_3 \!=\! \sigma_3/m \!>\! 0$ in Dirac spectrum of a UV-regularized 
theory such that the spatial IR dimension of Dirac eigenmodes at 
$t \!>\! t_3$ is the ordinary $d_\fir \!=\! 3$, while $d_\fir\!=\!0$ for 
modes at $0 \!<\! t \!<\! t_3$
\cite{Alexandru:2021pap, Alexandru:2021xoi, Alexandru:2023xho}. 
With that, $t_\text{ref} > t_3(a)$ for all sufficiently small $a$.
IR-Bulk decoupling in IR phase then may be formulated as follows. 
In IR phase there exists at least one $0 \!<\! \tilde{t} \!<\! t_\text{ref}$ 
such that the corresponding partitions of $\cF^2(0, t_\text{ref})$ 
decorrelate, and the decoupling scale $t_\text{dc}$ is the largest 
of such scales, namely
\begin{equation}
    \lim_{a \to 0} \lim_{L \to \infty} \;
    \langle \, \cF^2(0,\tilde{t}) \, \cF^2(\tilde{t}, t_\text{ref}) \, \rangle_\text{nc}
    \,=\, 0 
    \quad\quad \text{and} \quad\quad
    t_\text{dc} \,\equiv\, \sup \{\, \tilde{t} \,\}
    \quad
    \label{eq:141}            
\end{equation}
Note that segment $(\tilde{t}, t_\text{ref})$ contains part of bulk for 
each $\tilde{t}$.

Few remarks are important here. 
{\em (i)} It is implicitly understood that the above definition doesn't 
depend on the choice of $t_\text{ref}$. 
{\em (ii)} Regarding $t_\text{dc}$, the available 
evidence~\cite{Alexandru:2021pap, Alexandru:2021xoi, Meng:2023nxf,
Alexandru:2023xho} embodied in the metal-to-critical 
scenario~\cite{Alexandru:2021xoi} favors the simplest 
possibility that $t_\text{dc} \!=\! t_3 \!=\! t_A$ (see also Ref.~\cite{Horvath:2025ypt}), 
where $t_A \!>\! 0$ is the Anderson-like point of 
Refs~\cite{GarciaGarcia:2006gr, Kovacs:2010wx, Giordano:2013taa}.  
However, studies directly confirming the above form of IR-Bulk decoupling, 
as well as direct computation of $t_\text{dc}$ are yet to be performed.
{\em (iii)} It is consistent with the current knowledge that decoupling
Eq.~\eqref{eq:141} may occur already before taking the continuum 
limit, i.e. even at non-zero $a$. {\em (iv)} Detection of IR phase
(or even its definition) via the presence of the above IR-Bulk 
decoupling may be a fruitful approach to its numerical investigation.
{\em (v)} Normalized Pearson correlations of $\cF^2$ segments,
such as those entering Eq.~\eqref{eq:141} can be expressed via 
spectral values $\sigma_i(U)$ alone. The general expression is
\begin{equation}
    \big\langle \, \cF^2(\eta_1,\eta_2) \, \cF^2(\gamma_1, \gamma_2) 
    \, \big\rangle_\text{nc}
    \,=\, \Big\langle\,  
     \sum_{\eta_1 < \sigma_i \le \eta_2}  \, \sigma_i^2(U) \,
     \sum_{\gamma_1 < \sigma_j \le \gamma_2}  \, \sigma_j^2(U) \,      
     \, \Big\rangle_\text{nc}           
    \quad
    \label{eq:151}            
\end{equation}
as follows from Eqs.~\eqref{eq:121} and \eqref{eq:131}, or 
in the same way via spectral $t$-values.  

\smallskip
\noindent
{\bf 7.~The Uses: Coherent Lattice QCD.}
The term {\em coherent lattice QCD} refers to formulations of 
lattice-regularized QCD constructed from a single object, namely 
a suitable lattice Dirac operator. Such theories were first proposed 
in Refs.~\cite{Horvath:2006az, Horvath:2006md} and their construction 
utilized the same ideas as those employed here. The formulas involving
Dirac spectral density, suggested in this work, provide an additional 
insight in their construction that we now make explicit.

Let $D$ be any lattice Dirac operator with $\cs \!\neq\! 0$. The basic 
coherent lattice QCD has the form
\begin{equation}
     S \,=\, \frac{1}{2g^2} \, \frac{a}{\cs} \,
     \Trb \Bigl[ D(U) - D(\identity) \Bigr] 
     \,+\, a^4  \sum_{q=1}^{N_f} \psibar_q \Bigl( D(U) + m_q  \Bigr) \psi_q
     \label{eq:161}                 
\end{equation}
where $N_f$ is the number of quark flavors with masses $m_q$. In 
original Refs.~\cite{Horvath:2006az, Horvath:2006md} this action 
was written in terms of dimensionless lattice objects and factors involving
powers of $a$ were thus absent. Note that the gauge part is in fact 
$\cF^2(U)/2g^2$ with $\cF^2$ given in Eq.~\eqref{eq:111}.

The free-field part of the glue action is a constant that can be omitted 
in the definition of the theory. The effective glue action after integrating 
out the quark variables reads
%Using lattice units in $\hat{D}(U) = a D(U)$ and $\hat{m}_f = a m_f$ we have 
%\begin{equation}
%     S_\text{eff} \,=\, S_\text{eff}(U) \,=\,  
%     \frac{1}{2g^2} \, \frac{1}{\cs} \, \Trb \, \hat{D} \,-\, 
%     \sum_{f=1}^{N_f} \, \Trb\, \ln \bigl( \hat{D} + \hat{m}_f \bigr)  
%     \quad
%\end{equation} 
\begin{equation}
\begin{split}
     S_\text{eff}(U) &\,=\, \frac{1}{2g^2} \, \frac{a}{\cs} \, \Trb \, D(U) 
     \,-\,  \sum_{q=1}^{N_f} \, \Trb\, \ln \big( aD(U) + a m_q \big)  
     \, = \, \\ 
     &\,=\, V_4 \int_{\R^2[\C]} d\cS \, \rho_s(\lambda,U) \,
     \Bigl[\,  \frac{1}{2g^2} \frac{a}{\cs} \, \lambda \,-\,
     \sum_{q=1}^{N_f} \ln ( a \lambda + a m_q) \,\Bigr]
\end{split}
     \label{eq:171}                      
\end{equation} 
where $\lambda$ is a complex variable and the notation was introduced 
in connection with Eq.~\eqref{eq:017}. Thus, the defining object of 
the theory, the action, is scale-decomposed  and expressed in terms 
of Dirac spectral density. There is another hidden constant in the above 
expression, isolated by factoring $am_q$ in log terms. Discarding this 
constant leads to the effective action in the form
\begin{equation}
     S_\text{eff}(U) \,=\, 
     V_4 \int_{\R^2[\C]} d\cS \, \rho_s(\lambda,U) \,
     \Bigl[\,  \frac{1}{2g^2} \frac{a}{\cs} \, \lambda \,-\,
     \sum_{q=1}^{N_f} \ln \big( 1 + \lambda/m_q \big) \,\Bigr]
     \, \equiv \, S_\text{eff}^{\scriptscriptstyle{G}}(U) + 
     S_\text{eff}^{\scriptscriptstyle{Q}}(U)
     \label{eq:181}                      
\end{equation} 

For the family of overlap Dirac operators we then have in particular for
glue and quark parts
\begin{equation}
     2g^2 \, S_\text{eff}^{\scriptscriptstyle{G}}(U) \,=\,
     \frac{2\Delta}{\cs} \, n_0(U) \,+\, 
     \frac{a^2 V_4}{\cs \Delta}
     \int_{0}^{\bigl(\frac{2\Delta}{a}\bigr)^-} \! d\sigma \, 
     \rho^\eff(\sigma, U)  \, \sigma^2  
     \label{eq:191}                      
\end{equation} 
\begin{equation}
     -S_\text{eff}^{\scriptscriptstyle{Q}}(U) \,=\,
     n_0(U) \sum_{q=1}^{N_f} \ln \Bigl( 1 \,+\, \frac{2\Delta}{am_q} \Bigr) 
     \,+\,  V_4 \, \int_{0^+}^{\bigl(\frac{2\Delta}{a}\bigr)^-} \! d\sigma \, 
      \rho^\eff(\sigma, U) \, 
      \sum_{q=1}^{N_f}\,
      \ln \Biggl[ 1 + \frac{\sigma^2}{m_q^2} \Bigl(1 + am_q/\Delta \Bigr) \Biggr]  
      \label{eq:201}                      
\end{equation} 
where the contribution of real modes was again separated. Construction of other 
coherent lattice QCD actions, such as the symmetric logarithmic 
case~\cite{Horvath:2006az, Horvath:2006md}, proceeds along the same lines.  
 
\smallskip
\noindent
{\bf 8.~Epilogue.}
Dirac spectral density $\rho(\lambda)$ in QCD specifies 
the distribution of its quark degrees of freedom over Dirac scales. Its 
knowledge thus facilitates, among other things, our understanding of how 
these scales contribute to the composition of important quark observables 
such as $\langle \bar{\psi} \psi \rangle$. 
For example, we learned that $\langle \bar{\psi} \psi \rangle$ is a strictly IR 
quantity in the chiral limit, which is a particular way to interpret the approach 
of Banks and Casher~\cite{Banks:1979yr}.

Here we showed, via Eqs.~\eqref{eq:017} and \eqref{eq:051}, that 
$\rho(\lambda)$ also determines how scalar {\em glue density} gets 
aportioned across these scales. The new formulas reveal that glue density, 
in a stark contrast to quark density, is a strictly UV quantity. As such, it also 
provides for the analogue of Banks-Casher relation for gluon condensate. 
Details will be discussed in a dedicated account, but our reasoning makes 
it clear that this quantity is encoded in $1/\lambda$ power term in UV 
asymptotics of $\rho(\lambda)$.

The outlook on $\rho(\lambda)$ as the distribution of DOFs over scales 
was pervasive in works that led to the discovery of IR 
phase~\cite{Alexandru:2015fxa, Alexandru:2019gdm, Alexandru:2021pap, 
Alexandru:2021xoi}. Definition of the phase is also expressed via IR 
asymptotics of $\rho(\lambda)$~\cite{Alexandru:2019gdm}.
%\eqref{eq:091}
But its key features, such as scale invariance and IR-Bulk separation, 
were always chiefly attributed to gluonic rather than quark degrees of 
freedom. This was not contradictory with the explanation that matrix 
elements of the Dirac operator are in fact gauge-covariant glue 
operators, which makes $\rho(\lambda)$ the unusual (scale-dependent) 
{\em glue operator} as well. Here we made the needed direct link 
between Dirac spectral density and the aforementioned driving glue effects. 

Our formulas have several applications, particularly in relation to IR phase 
where these considerations first arose. Here we discussed the question of 
glue scale invariance of the IR part and the precise definition of IR-Bulk 
decoupling. They are all based on our formulas featuring Dirac spectral 
density, which possibly also opens new ways of numerical evaluation. Our 
considerations are most potent in regularizations known as coherent lattice 
QCD~\cite{Horvath:2006az, Horvath:2006md}, whose full effective actions 
can also be expressed via Dirac spectral density. 
 
Finally, we wish to emphasize that although the use of overlap Dirac 
operator for expressing gluon condensate (Eq.~\eqref{eq:051}) has many 
theoretical advantages, simpler operators may be preferable in numerical 
studies. In particular, the formula for all lattice Dirac operators with 
$\cs \!\neq\! 0$ is given in~\eqref{eq:017}. For example, simple Wilson-like 
Dirac operators with $\cs \!\neq\! 0$ can be considered.

\vfill\eject

\noindent
{\bf Acknowledgments.} 
Discussions with Vladim\'ir Balek and Robert Mendris on the topics of this 
work are gratefully acknowledged. Kudos to Hali for her gentle support.  
Many thanks to Dimitris Petrellis for help with technical aspects. 
I may never be able to repay my debts to Sylvia and Vlado. 
This work was supported in part by US Department of Energy under grant
DE-FG02-95ER40907.

%\vfill\eject

\bibliographystyle{JHEP}
\bibliography{my-references}

\end{document}